\documentclass[runningheads]{llncs}
\usepackage[T1]{fontenc}
\usepackage[utf8]{inputenc}
\usepackage{graphicx}
\usepackage{booktabs}
\usepackage{longtable}
\usepackage{array}
\usepackage{calc}
\usepackage{amsmath,amssymb}
\usepackage{textcomp}
\usepackage{algorithm}
\usepackage{algpseudocode}
\usepackage{etoolbox}
\usepackage{url}
\usepackage{microtype}
\usepackage[hidelinks]{hyperref}
\AtBeginEnvironment{longtable}{\small}
\AtBeginEnvironment{verbatim}{\small}
\providecommand{\real}[1]{#1}
\providecommand{\tightlist}{\setlength{\itemsep}{0pt}\setlength{\parskip}{0pt}}

\begin{document}
\title{O-Funnel: Lossless Structural Capture and Requirement-Driven Extraction from Drifting, Heterogeneous Documents}
\titlerunning{O-Funnel}
\author{Osama Mustafa}
\authorrunning{O. Mustafa}
\institute{Intellusion\\ \email{osama@intellusionai.com}}
\maketitle

\begin{abstract}
Pulling a fixed set of fields out of documents that arrive in many
formats and under drifting schemas is usually done with hand-written
byte patterns, which break whenever a key is renamed, a value is
reformatted, or a lookalike value appears first. We argue the cause is
structural: one pattern must both describe the value and locate it among
its surroundings. O-Funnel separates the two. It transcribes any XML,
JSON, CSV, HTML or key-value text document into one typed tree over five
constructors, gated by an oracle that rejects any capture that does not
reconstruct its source. Each needed field is declared in the tree's own
terms and located by fusing independent evidence (key, path, value
shape, synonym, key spelling, record neighborhood, value profile), so
the best-supported node wins and a missing field is reported with a
reason. Data no requirement claims becomes residue that a funnel traces
back to the requirements to learn new key aliases. On 34,989 real PubMed
records, O-Funnel matches a hand-written parser (F1 1.00). After a
five-element schema rename, the parser's regular expressions fall to
0.20 while O-Funnel stays at 1.00, with every capture verified complete.
On constructed suites that isolate regex failure modes it raises F1 from
0.43 to 1.00, and from 0.80 to 0.94 after self-improvement; on held-out
schema-matching instances it is competitive with classical matchers
without training. O-Funnel is a dependency-free Python library (pip
install ofunnel).
\keywords{information extraction \and semi-structured data \and schema drift \and schema matching \and data integration \and provenance \and self-supervised alias discovery}
\end{abstract}

\section{Introduction}\label{introduction}

A great deal of practical software does nothing more glamorous than turn
documents into records: given a stream of files, pull out the handful of
fields a downstream system needs. When every file has the same shape
this is trivial. In reality the files do not have the same shape. The
same logical field appears under different key spellings
(\texttt{received\_date}, \texttt{dateReceived}, \texttt{RECEIVED\_DT}),
in different formats (an XML element this week, a JSON object next week,
a CSV export after that), with different value encodings
(\texttt{2026-06-17}, \texttt{June\ 17,\ 2026}, \texttt{06/17/2026}),
next to decoys of the same shape (a related identifier printed before
the real one), and sometimes under keys that carry no meaning at all
(\texttt{c1}, \texttt{f7}, column 4). Each variation is easy to handle
alone; the difficulty is that they arrive together, unannounced, and
keep changing.

The reflexive answer is the regular expression. A regex is a compact,
fast, widely understood way to say ``a value that looks like this.'' But
it operates on the byte stream, and the byte stream mixes the value with
everything around it. To find a value, a regex must encode in one
pattern an assumption about the value and an assumption about its
surroundings, because it has no other way to tell occurrences apart.
When either assumption breaks, the pattern fails: it silently returns
nothing, or, worse, confidently returns the wrong span. This is not a
defect of one pattern that a better pattern would fix; it is a property
of matching a structured document through an unstructured view of it.

The problem sits at the boundary between two spaces. The
\textbf{document space} is an open set of formats and schemas that we do
not control and cannot enumerate. The \textbf{requirement space} is the
small, fixed set of things we need, which we do control. A regex tries
to be fluent in both at once and ends up fluent in neither: the document
space is too varied for one pattern, and the requirement is written in
the wrong vocabulary (bytes) for its intent (a field).

O-Funnel separates the two spaces. It first transcribes any document,
losslessly, into one structural representation, so the document space
collapses to a single tree whatever its dialect. It then states each
requirement in the language of that tree and resolves it using many
independent kinds of evidence at once. Nothing is matched against raw
bytes; everything is matched against structure, keys, types, value
shapes, neighbors and value populations, fused into one score. A
requirement that cannot be located is reported as a provable absence
with a reason, never as silence. The stance is also inverted: rather
than only hunting for what was declared, O-Funnel keeps an open space
for whatever else the document carries. Every captured item no
requirement claimed becomes \textbf{residue}, and the residue is the
signal for what to learn next. A funnel traces it back to the
requirements and proposes new key aliases, so coverage grows as the
system sees more drift instead of decaying. The system is named after
that funnel.

This paper presents O-Funnel as a solution and shows how it behaves in
challenging conditions. It is not a leaderboard study. Our contributions
are:

\begin{enumerate}
\def\labelenumi{\arabic{enumi}.}
\tightlist
\item
  \textbf{Lossless structural capture} from five common formats into one
  typed, labeled, ordered tree over a closed vocabulary of five
  constructors, with a completeness oracle that gates every capture
  (Section 4.1).
\item
  \textbf{A declarative requirement language and a fused evidence
  ladder} (Sections 4.2 to 4.3): candidates reached by several
  independent rungs merge, agreeing evidence adds bounded confidence,
  and a contradicted declaration subtracts, so the winner is the
  best-supported node rather than the first one found.
\item
  \textbf{Domain-free witnesses}: a value profile derived from a few
  example values, and a key-spelling similarity that understands
  abbreviations, stems, prefixes and initialisms (Sections 4.4 and 4.6).
\item
  \textbf{The residue funnel}, a structure-based self-improvement loop
  that turns unclaimed data into alias proposals and promotes only the
  safe ones, and a \textbf{scavenger} that recovers fields under opaque
  keys but abstains on ambiguity (Sections 4.5 and 4.7).
\item
  \textbf{Demonstrations in challenging conditions}: 34,989 real PubMed
  records before and after a realistic schema drift, constructed suites
  that isolate the failure modes of byte-level extraction, a
  self-improvement study, and a schema-matching sanity check, all
  reproducible with released scripts (Section 6).
\end{enumerate}

Section 4.1 builds the capture (Figure 2), Sections 4.2 to 4.8 resolve
requirements over it (Figure 1), Section 6 demonstrates it, and Section
7 discusses implications and limits. The library and every experiment
are public (Section 9).

\begin{figure}
\centering
\includegraphics[width=\linewidth,height=\textheight,keepaspectratio,alt={O-Funnel architecture. A document in any supported format is turned into one structural tree by a per-format adapter and an additive normalization pass, then gated by a completeness oracle. Requirements, declared in the same structural language, are resolved against the tree by a fused evidence ladder that returns matches (with method, confidence and provenance), provable absences, and a residue of unclaimed data. The residue feeds the O-Funnel, which proposes and promotes new key aliases back into the requirements.}]{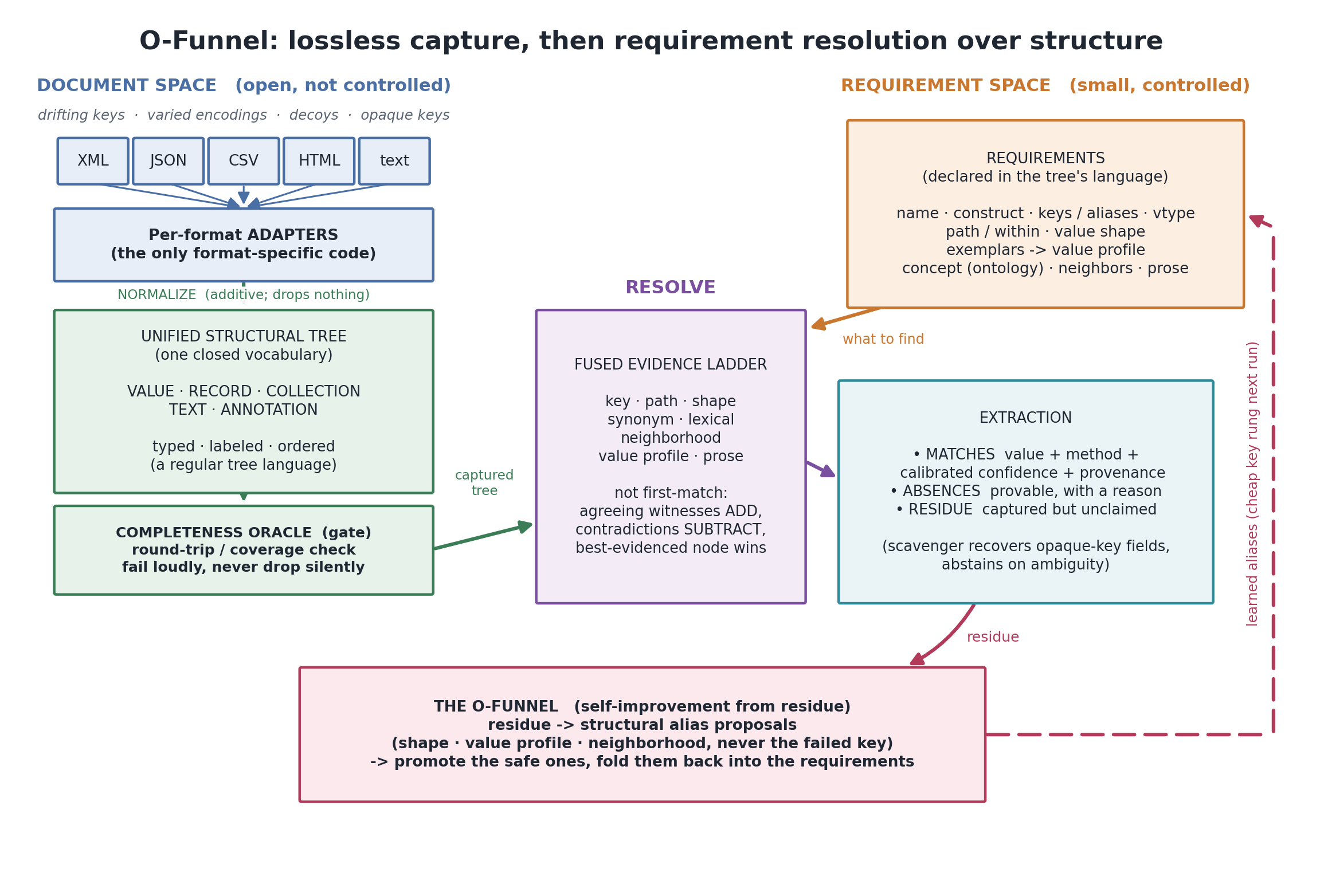}
\caption{\textbf{O-Funnel architecture.} A document in any
supported format is turned into one structural tree by a per-format
adapter and an additive normalization pass, then gated by a completeness
oracle. Requirements, declared in the same structural language, are
resolved against the tree by a fused evidence ladder that returns
matches (with method, confidence and provenance), provable absences, and
a residue of unclaimed data. The residue feeds the O-Funnel, which
proposes and promotes new key aliases back into the requirements.}
\end{figure}

\section{Related Work}\label{related-work}

\textbf{Rule-based extraction.} Hand-written patterns remain the
workhorse of production extraction, and Chiticariu et al.~{[}1{]} argue
that rule systems are underserved by research rather than obsolete. Our
position is compatible: we keep declarative rules, but move them off the
byte stream and onto structure, and add automatic repair. Sarawagi's
survey {[}2{]} frames the broader information-extraction landscape.

\textbf{Wrapper induction and web extraction.} Wrapper induction
{[}3{]}, RoadRunner's grammar inference {[}4{]}, the SWDE benchmark and
its unified solution {[}5{]}, and DIADEM {[}6{]} learn extractors that
exploit the repeated template of a single site. O-Funnel is
complementary: it assumes no shared template, only that each document is
well formed in a supported format, and it targets a fixed set of
requirements rather than every record on a page.

\textbf{Schema matching.} Aligning attributes across schemas is surveyed
by Rahm and Bernstein {[}7{]}. Classical matchers include Cupid {[}8{]},
Similarity Flooding {[}9{]} and the COMA family {[}10, 11{]}, and
learning-based reconciliation such as LSD {[}12{]}; the Valentine study
{[}13{]} provides a common corpus and protocol, which we reuse. More
recent matchers are learned or language-model based: Unicorn {[}27{]}
trains one model for many matching tasks, and Magneto {[}28{]} and
ReMatch {[}29{]} use large language models with retrieval. Our
resolution ladder is a one-sided instance-and-schema matcher that aligns
captured keys to a fixed requirement set, with fallbacks to value shape,
neighborhood and value population when names fail. It needs no training
data or model.

\textbf{Language models over heterogeneous data.} Foundation models can
perform many data-wrangling tasks from prompts {[}25{]}, and Evaporate
{[}26{]} builds structured views of heterogeneous semi-structured
documents by having a language model synthesize extraction code. These
systems gain breadth and semantic reach; O-Funnel instead offers a
deterministic, auditable, dependency-free resolver whose every value
carries its method and evidence, and which can serve as a cheap first
stage or a baseline for them.

\textbf{Semi-structured data models.} One self-describing model for
heterogeneous sources goes back to OEM {[}14{]} and the semi-structured
data program of Abiteboul {[}15{]} and Buneman {[}16{]}; web tables
{[}23{]} are a related line for tabular data, and tree automata {[}17{]}
give the formal background for labeled ordered trees. Our tree is in
that lineage. What we add is a closed five-constructor vocabulary, an
explicit completeness oracle, and a resolution layer written in the same
vocabulary.

\textbf{Weak supervision and calibration.} Snorkel {[}18{]} combines
noisy labeling functions and distant supervision {[}19{]} bootstraps
extractors from a knowledge base. Our funnel is a structural analogue
that proposes key aliases from unclaimed data using agreement between
structural signals. The calibration hook follows the
empirical-reliability view of Guo et al.~{[}20{]}. String-similarity
components draw on edit distance {[}21{]}, the Ratcliff/Obershelp
matcher {[}22{]}, and the comparison of name-matching metrics by Cohen
et al.~{[}24{]}.

\section{Problem Formulation}\label{problem-formulation}

Let a \textbf{document} \emph{d} be a byte string in one of a set of
supported formats \emph{F}. Let a \textbf{requirement set}
\(R = \{r_1, \ldots, r_k\}\) be a small, fixed collection of fields to
extract, each with whatever is known about it: candidate names, type,
value shape, a few example values, structural context. An
\textbf{extraction} is a partial map from \emph{R} to (value,
provenance, confidence) triples. We want an extractor \emph{E(d, R)}
that is:

\begin{itemize}
\tightlist
\item
  \textbf{complete or loud}: it never silently drops document content,
  and it reports which requirements it could not satisfy, and why;
\item
  \textbf{format-blind above capture}: the logic that resolves \emph{R}
  does not depend on which format produced \emph{d};
\item
  \textbf{drift-robust}: its accuracy degrades gracefully as key names,
  value encodings, formats and nearby decoys change;
\item
  \textbf{auditable}: every value carries how it was found and how
  confident the system is.
\end{itemize}

The regex baseline instantiates \emph{E} as one byte pattern per field.
It is fast, but a miss is silence, a decoy is a confident error, and a
renamed key or reformatted value breaks it. The two-space view makes
this precise: one expression must (a) pick out a value population and
(b) tell the target occurrence apart from other occurrences, and it has
only the byte neighborhood to do (b). O-Funnel gives (a) and (b)
separate machinery. Value description (shape, value profile) lives in
the requirement; localization (key, path, neighborhood) lives in the
captured tree.

\section{Method}\label{method}

O-Funnel has two halves. The first (Section 4.1) is per-format and turns
a document into one tree. The second (Sections 4.2 to 4.8) is
format-blind and resolves requirements against that tree.

\textbf{A running example.} Take one record with two fields: an
identifier whose values look like \texttt{4021-KP73} and a
\texttt{received} date. It may arrive as the XML, JSON or CSV of Figure
2; with its key renamed (\texttt{code} becomes \texttt{c1} or
\texttt{ref}); with the date reformatted (\texttt{June\ 17,\ 2026}); or
with a same-shaped decoy (\texttt{prior\_code}) placed before the real
value. No single byte pattern survives all of these. The sections below
show which part of O-Funnel handles each variation.

\subsection{Lossless structural
capture}\label{lossless-structural-capture}

\textbf{Adapters.} For each format a small adapter builds a tree of
nodes. A node carries a \texttt{key} with a \texttt{key\_source}
(\emph{named} for a JSON member or XML tag, \emph{indexed} for an array
position, \emph{inferred} for a label parsed from a text line), a
\texttt{kind} (scalar or group), an optional scalar \texttt{value},
ordered children, a \texttt{role} (element, attribute, text, comment)
and a byte span where the format provides one. The adapters are the only
format-specific code.

\textbf{The unified vocabulary.} A normalization pass classifies every
node into exactly one of five constructors, the closed vocabulary all
later code speaks (Table 1). A VALUE also carries a type (string,
number, boolean or null). Classification is additive: it sets the
construct, an \texttt{origin} recording the source dialect (for example
\texttt{xml.attribute}, \texttt{json.array}) and the type, and never
removes a node. The source dialect stays recoverable, yet everything
downstream can ignore it. Formally, a captured document is a finite
labeled, ordered tree over these five constructors; we use the term
``language'' in this descriptive sense.

\begin{longtable}[]{@{}
  >{\raggedright\arraybackslash}p{(\linewidth - 4\tabcolsep) * \real{0.2273}}
  >{\raggedright\arraybackslash}p{(\linewidth - 4\tabcolsep) * \real{0.3409}}
  >{\raggedright\arraybackslash}p{(\linewidth - 4\tabcolsep) * \real{0.4318}}@{}}
\caption{The five constructors of the unified
vocabulary.}\tabularnewline
\toprule\noalign{}
\begin{minipage}[b]{\linewidth}\raggedright
constructor
\end{minipage} & \begin{minipage}[b]{\linewidth}\raggedright
meaning
\end{minipage} & \begin{minipage}[b]{\linewidth}\raggedright
example sources
\end{minipage} \\*
\midrule\noalign{}
\endfirsthead
\toprule\noalign{}
\begin{minipage}[b]{\linewidth}\raggedright
constructor
\end{minipage} & \begin{minipage}[b]{\linewidth}\raggedright
meaning
\end{minipage} & \begin{minipage}[b]{\linewidth}\raggedright
example sources
\end{minipage} \\*
\midrule\noalign{}
\endhead
\bottomrule\noalign{}
\endlastfoot
VALUE & an atomic typed value & JSON scalar, XML leaf element or
attribute, CSV cell \\*
RECORD & a keyed group of heterogeneous fields & JSON object, XML
element with children, CSV row \\*
COLLECTION & an ordered group of like items & JSON array, CSV table \\*
TEXT & prose or inline text & XML text node, a text line \\*
ANNOTATION & non-content markup & XML comment, processing instruction \\*
\end{longtable}

\textbf{Text-bearing elements.} Markup formats contain elements that
carry their own text without being leaves: \emph{simple content}, an
element with attributes and text
(\texttt{\textless{}PMID\ Version="1"\textgreater{}41605291\textless{}/PMID\textgreater{}}),
and \emph{mixed content}, text with inline markup
(\texttt{\textless{}ArticleTitle\textgreater{}Effect\ of\ \textless{}i\textgreater{}X\textless{}/i\textgreater{}\ on\ Y\textless{}/ArticleTitle\textgreater{}}).
Capture keeps them losslessly as a RECORD with attribute and text
children. Resolution reads their text through a value view, so a
requirement for a value reaches them; Section 6.2 shows why this matters
on real data.

\textbf{The completeness oracle.} Every capture is gated. For XML, JSON
and CSV, whose structure is self-describing, the oracle independently
re-parses the source and checks that both trees contain the same content
in the same order, ignoring only insignificant whitespace and honoring
the declared encoding. For key-value text it checks byte coverage: every
source byte belongs to some leaf. HTML is presentational and often
malformed, so its guarantee is weaker: the oracle checks that no visible
text is dropped (content-lossless), not that every attribute and
structural detail round-trips. If the oracle fails, the run fails; a
capture that cannot show it preserved its input is never used.

\begin{algorithm}[t]
\caption{CAPTURE$(raw, fmt)$}
\begin{algorithmic}[1]
\State $fmt \gets fmt$ \textbf{or} \Call{Sniff}{$raw$}
\State $tree \gets \Call{Adapter}{fmt, raw}$ \Comment{format-specific: builds the node tree}
\State \Call{Normalize}{$tree, fmt$} \Comment{additive: sets construct, origin, type}
\If{\textbf{not} \Call{CheckComplete}{$raw, tree, fmt$}}
  \State \textbf{fail} \Comment{re-parse or coverage gate}
\EndIf
\State \Return $(tree, fmt)$
\end{algorithmic}
\end{algorithm}

\begin{figure}
\centering
\includegraphics[width=\linewidth,height=\textheight,keepaspectratio,alt={Lossless capture. The same record in three dialects (XML, JSON, CSV) is transcribed into one structural tree. Keys, values and order are identical across dialects; the source dialect of each node is kept in origin, so nothing is lost. An independent re-parse (the oracle) must reconstruct the source or the run fails.}]{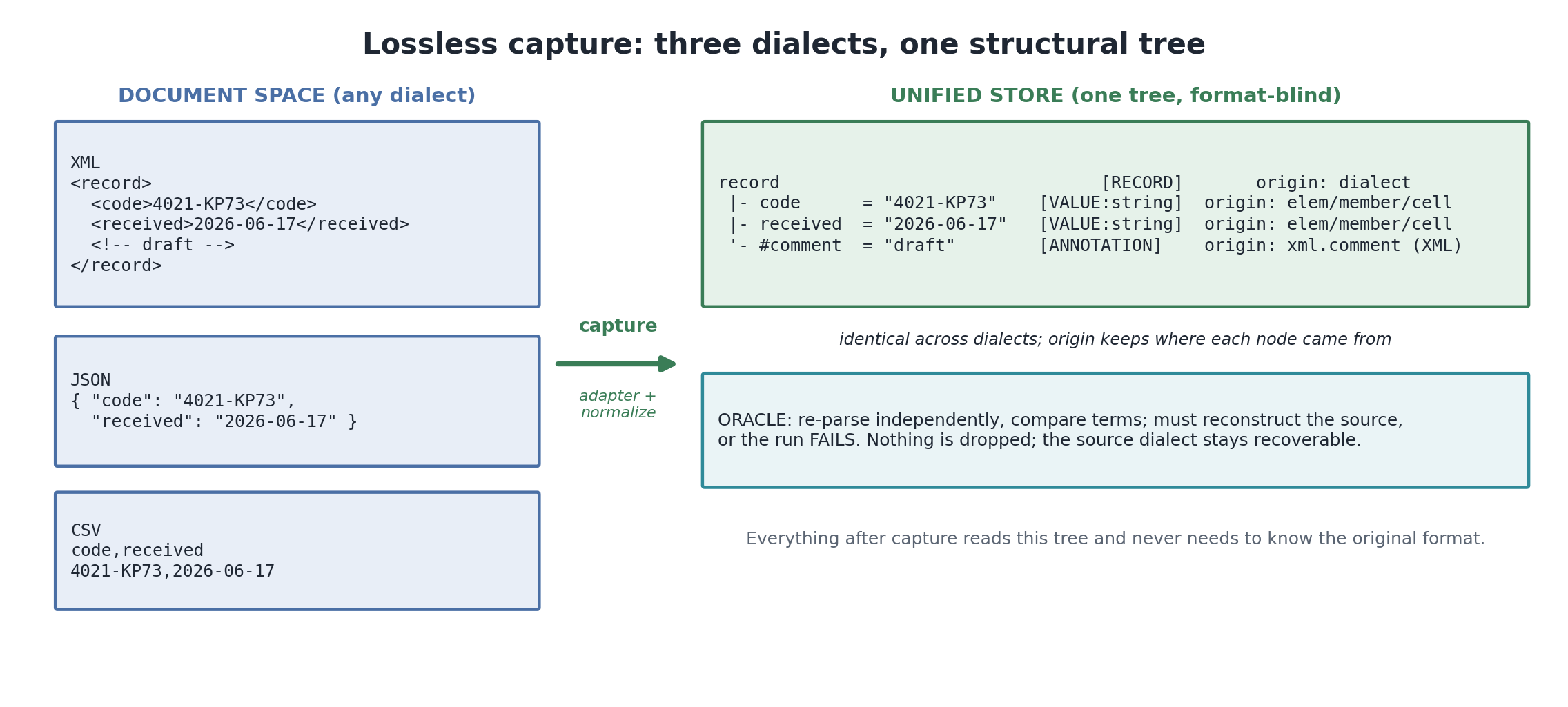}
\caption{\textbf{Lossless capture.} The same record in three
dialects (XML, JSON, CSV) is transcribed into one structural tree. Keys,
values and order are identical across dialects; the source dialect of
each node is kept in \texttt{origin}, so nothing is lost. An independent
re-parse (the oracle) must reconstruct the source or the run fails.}
\end{figure}

\subsection{The requirement language}\label{the-requirement-language}

A requirement is declared in the tree's vocabulary, not as a byte
pattern:

\begin{verbatim}
Requirement(name, construct=VALUE,
    keys=(alias, ...),       # key spellings (case/separator free)
    vtype=(...),             # allowed value types
    path=(...), within=...,  # containment context
    shape=regex,             # value shape, tested on one value
    exemplars=(...),         # example values -> value profile
    concept=...,             # ontology concept (synonyms)
    neighbors=(...),         # sibling keys in the same record
    prose=regex,             # phrase, only inside prose nodes
    normalize=...)           # bool, number, date or text
\end{verbatim}

The value \texttt{shape} is still a regular expression, but a much tamer
one. It is matched against the content of a single node that has already
been located, never against the document. It describes the value only;
localization is handled structurally. This is what defuses decoys: when
any structural cue is declared, a shape can confirm a candidate but
cannot by itself choose one occurrence over another.

\subsection{Resolution as fused
evidence}\label{resolution-as-fused-evidence}

Resolving a requirement walks the tree once and gathers, for each
candidate node, every rung of evidence that fires (Table 2).

\begin{longtable}[]{@{}
  >{\raggedright\arraybackslash}p{(\linewidth - 4\tabcolsep) * \real{0.2093}}
  >{\raggedright\arraybackslash}p{(\linewidth - 4\tabcolsep) * \real{0.4884}}
  >{\raggedright\arraybackslash}p{(\linewidth - 4\tabcolsep) * \real{0.3023}}@{}}
\caption{The rungs of the evidence ladder.}\tabularnewline
\toprule\noalign{}
\begin{minipage}[b]{\linewidth}\raggedright
rung
\end{minipage} & \begin{minipage}[b]{\linewidth}\raggedright
locates or confirms by
\end{minipage} & \begin{minipage}[b]{\linewidth}\raggedright
nominal confidence
\end{minipage} \\*
\midrule\noalign{}
\endfirsthead
\toprule\noalign{}
\begin{minipage}[b]{\linewidth}\raggedright
rung
\end{minipage} & \begin{minipage}[b]{\linewidth}\raggedright
locates or confirms by
\end{minipage} & \begin{minipage}[b]{\linewidth}\raggedright
nominal confidence
\end{minipage} \\*
\midrule\noalign{}
\endhead
\bottomrule\noalign{}
\endlastfoot
key & the node's key equals an alias (element / attribute / inferred
key) & 0.95 / 0.90 / 0.75 \\*
path & the key path ends with the declared containment & 0.90 \\*
shape & the value matches \texttt{shape}; locates only when no
structural cue is declared & 0.70 \\*
synonym & the key is a registered surface form of \texttt{concept} &
0.90 exact, up to 0.85 fuzzy \\*
lexical & the key is the same words spelled differently & 0.55 + 0.30 x
similarity, up to 0.85 \\*
neighborhood & the sibling that fits, inside a record holding the
declared neighbors & 0.65, partial credit below \\*
value & the value fits the requirement's value profile & witness only \\*
prose & the phrase, inside prose nodes only & 0.50 \\*
absent & nothing represents the field; a reason is recorded & value
None \\*
\end{longtable}

\textbf{Fusion, not first match.} A clean key or path hit with
confidence at least 0.9 is a fast path; it is the common, pristine case
and runs at near key-lookup speed. Otherwise every rung's candidates are
gathered and merged per node. Let \(K(n)\) be the set of rungs that
reached node \(n\) and \(s_k(n)\) the confidence rung \(k\) assigns
after its own witnesses. The fused score is

\[ S(n) = \min\Big(c,\ \max_{k \in K(n)} s_k(n) + \alpha\,(|K(n)| - 1)\Big), \]

with \(\alpha = 0.04\) and cap \(c = 0.95\) (0.99 for a clean structural
hit), and the node with the highest \(S\) wins, ties broken by document
order. Within each rung, an agreeing witness (a matching shape, or a
value that fits the profile) adds \(\alpha\); a contradicted declaration
(a located value that fails the declared shape or clearly misfits the
profile) multiplies the score by \(\beta = 0.8\) and is tagged
\texttt{shape\_miss} or \texttt{value\_miss} rather than discarded; a
value that fails its declared normalizer is multiplied by 0.6. A field
found by a fuzzy key spelling, among the right neighbors and with the
right value population is therefore trusted more than any one of those
signals alone. A field whose key matches but whose value contradicts the
declared shape is still returned, flagged and down-weighted, so a
reviewer can see the tension. The constants are hand-set defaults; we
did not tune them per experiment, and all experiments use the same
values.

\begin{algorithm}[t]
\caption{RESOLVE$(tree, req, synonyms)$}
\begin{algorithmic}[1]
\State $top \gets \Call{Structural}{tree, req}$ \Comment{key, path, shape}
\If{the best of $top$ is a key or path hit with confidence $\ge 0.9$}
  \State \Return it \Comment{fast path}
\EndIf
\State $G \gets [\,top\,]$
\If{$req.concept$} \State $G \gets G + \Call{Synonym}{tree, req, synonyms[req.concept]}$ \EndIf
\If{$req.keys$} \State $G \gets G + \Call{Lexical}{tree, req}$ \EndIf
\If{$req.neighbors$} \State $G \gets G + \Call{Neighborhood}{tree, req}$ \EndIf
\State $F \gets \Call{Fuse}{G}$ \Comment{merge per node, score $S(n)$}
\If{$F \neq \emptyset$} \State \Return $\arg\max_{n} S(n)$ \EndIf
\If{$req.prose$} \State \Return \Call{Prose}{$tree, req$} \EndIf
\State \Return \Call{Absent}{$reason$} \Comment{summary of rejections}
\end{algorithmic}
\end{algorithm}

\textbf{Absence is a result.} When nothing fires, the requirement is
reported absent with a machine-readable reason (\texttt{no\_candidate},
\texttt{shape\_rejected:n}, \texttt{vtype\_rejected:n},
\texttt{ambiguous:c1,c2}). Silence is never an outcome.

\subsection{Value profiles}\label{value-profiles}

Names lie; value populations rarely do. From a requirement's
\texttt{exemplars} O-Funnel derives a profile: for each example a
run-length character-class signature (\texttt{4021-KP73} becomes
\texttt{d4-A2d2}, coarsened to \texttt{d-Ad}), plus the mix of digits,
letters, spaces and punctuation, the mean length, and the fraction of
values that are numbers, dates or booleans. Two profiles are compared by
a cosine over signature distributions blended with the mix, length and
kind fractions. The result is a purely structural, domain-free witness.
Two fields with the same key but different value populations are
recognized as different, so a decoy sharing a name cannot capture a
requirement. Two fields with different keys but the same population are
recognized as probably the same, so a fully renamed key can still be
scored by what its values look like. The single-value form scores one
value against a profile (1.0 when its fine signature was seen, 0.85 for
a coarse match, a scaled similarity otherwise) and acts as a witness on
every rung, in the funnel and in the scavenger.

\subsection{The residue funnel: self-improvement from unclaimed
data}\label{the-residue-funnel-self-improvement-from-unclaimed-data}

After resolution, the captured values that no requirement claimed form
the \textbf{residue}. Across a corpus, the funnel asks of each residue
item which requirement it most likely feeds, judged only by structure:
value shape, value profile, neighborhood and nearness of its key to the
requirement's aliases, and never the key that already failed. An item
that fits a requirement is \emph{funnelled} to it as an alias proposal;
an item that fits nothing stays in the residue, an honest measure of
what is still unknown.

A proposal carries its support (how many items back it), its mean
confidence, and a \emph{contest} count: how often the same item also
fits another requirement, two items in one record fit the same
requirement, or the requirement was already satisfied in that record.
Contested proposals are the dangerous ones, because a wrong alias is not
a miss but corruption, so promotion keeps only proposals with enough
support, high confidence and a low contest ratio. Promoted aliases are
folded into the requirements' keys, so the next run resolves them on the
cheap key rung. A pipeline can run this continuously in windows.

Because the funnel uses the value profile both as a hard gate (an item
whose values contradict the profile is never funnelled) and as positive
evidence (an item whose values match is funnelled even when its key and
neighbors say nothing), it can trace free-text fields, which have no
value shape, by the stability of their value population alone (Section
6.4).

\begin{figure}
\centering
\includegraphics[width=\linewidth,height=\textheight,keepaspectratio,alt={The O-Funnel loop. Every captured item no requirement claimed enters the funnel and is scored per requirement on structural fit only (value shape, value profile, neighborhood, key nearness), never on the key that already failed; a contradicted value profile is a hard gate. Well-supported, uncontested fits become alias proposals, the safe ones are promoted into the requirements' keys, and the rest stay in the residue.}]{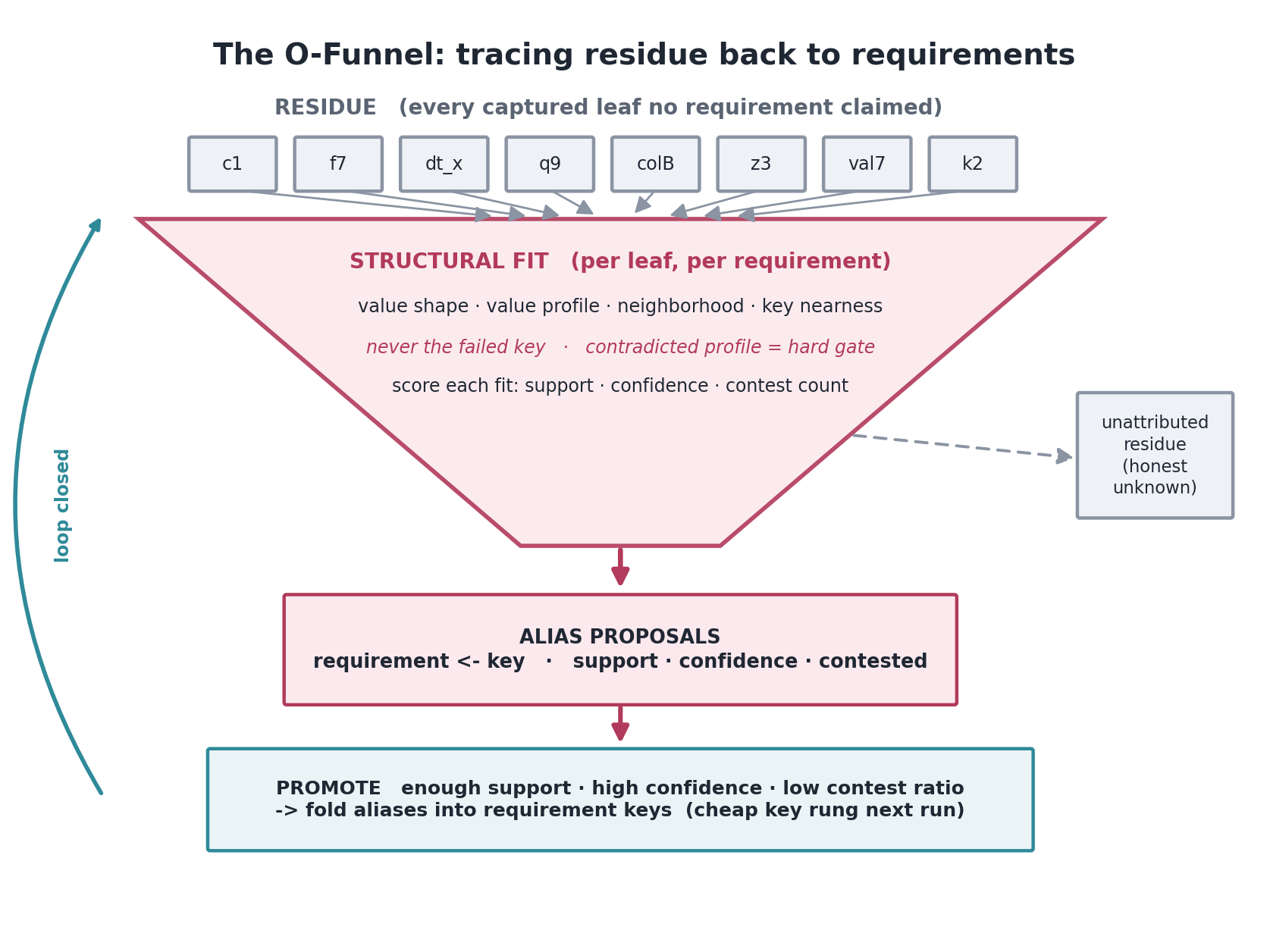}
\caption{\textbf{The O-Funnel loop.} Every captured item no
requirement claimed enters the funnel and is scored per requirement on
structural fit only (value shape, value profile, neighborhood, key
nearness), never on the key that already failed; a contradicted value
profile is a hard gate. Well-supported, uncontested fits become alias
proposals, the safe ones are promoted into the requirements' keys, and
the rest stay in the residue.}
\end{figure}

\subsection{Key-spelling similarity that knows how names are
abbreviated}\label{key-spelling-similarity-that-knows-how-names-are-abbreviated}

The lexical rung compares a captured key with a requirement's aliases
using a soft token-set similarity. Beyond exact token overlap it
credits, at 0.9, equivalences that plain edit distance misses: common
abbreviation groups (\texttt{dt} and \texttt{date}, \texttt{rcvd} and
\texttt{received}, \texttt{ref} and \texttt{reference}, \texttt{num} and
\texttt{number}), stems (\texttt{dates} and \texttt{date}), prefixes of
three or more characters (\texttt{descr} and \texttt{description}),
consonant-skeleton abbreviations whose short side has no vowels, and
initialisms (\texttt{pin} and
\texttt{personal\ identification\ number}). It deliberately does not
invent meaning: an acronym whose letters are not the initials of the
target scores low, and a short accidental edit-distance neighbor
(\texttt{title} and \texttt{total}) is not treated as equivalent.
Results are memoized; keys repeat across every record of a feed, so the
cost of this rung is paid once per distinct key.

\subsection{The scavenger: opaque keys without
guessing}\label{the-scavenger-opaque-keys-without-guessing}

When a requirement with a declared shape is left absent, the scavenger
searches the residue for values matching that shape. If exactly one
matches, it is claimed at low confidence (0.4) and tagged; this recovers
fields under fully opaque keys (\texttt{c1}, \texttt{f7}). If several
match, they are ranked only by evidence that can separate them (value
profile and neighborhood) and deliberately not by key similarity,
because a decoy such as \texttt{RELATED\_ID} is often the closest key. A
clear winner (margin at least 0.15) is claimed and tagged
\texttt{ranked}; otherwise the field stays absent with an
\texttt{ambiguous:} reason naming the candidates. These guards keep the
scavenger from turning back into a byte regex that grabs the first
plausible span.

\subsection{A calibration hook}\label{a-calibration-hook}

The confidences above are nominal. O-Funnel exposes a hook
(\texttt{reliability}) that returns the measured precision of a match's
(method, witnesses) bucket once the user has recorded it on labelled
data of their own, and the nominal confidence otherwise; it never raises
a contradicted or ambiguous match. We report no calibration result here,
since meaningful reliability must be measured on the deployment's data.

\section{Implementation}\label{implementation}

O-Funnel is released as a Python library of about 1,800 lines using only
the standard library: no third-party parsers, no machine-learning
frameworks, no network calls. Capture uses the standard XML, JSON and
CSV parsers plus a tolerant HTML parser and a small key-value text
reader; everything above capture is tree computation. The system is
deterministic: the same document and requirements always produce the
same extraction. The library ships 37 unit tests (capture and the oracle
across formats, each rung, fusion, value profiles, text-bearing
elements, the funnel, the scavenger), and a \texttt{benchmarks/}
directory holds the script behind every number in Section 6.

\textbf{Complexity.} Capture and the oracle are linear in document size.
Resolution walks the tree once per requirement, so a document with \(N\)
nodes and \(R\) requirements costs \(O(N \cdot R)\) node visits; the
non-key rungs add constant work per node, and key similarities are
memoized, so across a feed whose keys repeat they are effectively free
after first sight. The funnel is linear in the total residue of the
corpus.

\section{Evaluation}\label{evaluation}

The goal of this section is to show how O-Funnel behaves in conditions
where byte-level extraction is known to struggle, on a large real corpus
and on constructed scenarios, and to check its building blocks against
established tools. It is not a leaderboard study.

\subsection{Setup}\label{setup}

We compare against \textbf{byte regular expressions}, the incumbent this
work aims to replace. On the real corpus we also use an
\textbf{exact-path parser} written for the corpus schema, which serves
as gold. Unless noted, the metric is field-level F1 against gold values
after light normalization (whitespace collapsed, dates parsed). All runs
are single-threaded on one core with O-Funnel 0.1.1 and the same default
constants throughout. Synthetic examples use a generic identifier
pattern
(\texttt{\textbackslash{}d\{4\}-{[}A-Z{]}\{2\}\textbackslash{}d\{2\}}),
generic date encodings and neutral field names.

\subsection{Real-world corpus: PubMed before and after schema
drift}\label{real-world-corpus-pubmed-before-and-after-schema-drift}

We use 34,989 citation records from two files of the public PubMed 2026
baseline {[}30{]} (one early file with sparse historical records, one
recent file with rich records). Each
\texttt{\textless{}PubmedArticle\textgreater{}} is one document. Six
fields are requested: the PMID (an attribute-bearing element that also
recurs inside citation cross-references), the article title (315 titles
contain inline markup), the journal ISSN (attribute-bearing), the
language, the publication year (either a \texttt{Year} element or a
free-text \texttt{MedlineDate} from which the year is extracted, and
only inside \texttt{PubDate}, because several other date blocks also
contain years) and the DOI (held in generic elements such as
\texttt{\textless{}ELocationID\ EIdType="doi"\textgreater{}} and located
by value shape alone). Gold comes from an ElementTree parser that
addresses each field by its exact path. The regex baseline is what a
careful engineer would write against the same schema: tag patterns, with
inline markup stripped and entities decoded.

We then apply a realistic schema drift to every record, the kind a feed
introduces in a new version: \texttt{PMID} becomes
\texttt{PubMedIdentifier}, \texttt{ArticleTitle} becomes
\texttt{TitleOfArticle}, \texttt{ISSN} becomes \texttt{JournalISSN},
\texttt{Language} becomes \texttt{Lang}, and \texttt{PubDate} becomes
\texttt{PublicationDate}. The requirements are not changed. O-Funnel is
given a two-entry synonym map (surface terms for PMID and ISSN), the
kind of alias list a team maintains anyway.

\begin{longtable}[]{@{}
  >{\raggedright\arraybackslash}p{(\linewidth - 8\tabcolsep) * \real{0.1579}}
  >{\raggedleft\arraybackslash}p{(\linewidth - 8\tabcolsep) * \real{0.2105}}
  >{\raggedleft\arraybackslash}p{(\linewidth - 8\tabcolsep) * \real{0.2105}}
  >{\raggedleft\arraybackslash}p{(\linewidth - 8\tabcolsep) * \real{0.2105}}
  >{\raggedleft\arraybackslash}p{(\linewidth - 8\tabcolsep) * \real{0.2105}}@{}}
\caption{PubMed, 34,989 real records, F1 per field before and after a
five-element schema drift.}\tabularnewline
\toprule\noalign{}
\begin{minipage}[b]{\linewidth}\raggedright
field
\end{minipage} & \begin{minipage}[b]{\linewidth}\raggedleft
regex, pristine
\end{minipage} & \begin{minipage}[b]{\linewidth}\raggedleft
O-Funnel, pristine
\end{minipage} & \begin{minipage}[b]{\linewidth}\raggedleft
regex, drifted
\end{minipage} & \begin{minipage}[b]{\linewidth}\raggedleft
O-Funnel, drifted
\end{minipage} \\*
\midrule\noalign{}
\endfirsthead
\toprule\noalign{}
\begin{minipage}[b]{\linewidth}\raggedright
field
\end{minipage} & \begin{minipage}[b]{\linewidth}\raggedleft
regex, pristine
\end{minipage} & \begin{minipage}[b]{\linewidth}\raggedleft
O-Funnel, pristine
\end{minipage} & \begin{minipage}[b]{\linewidth}\raggedleft
regex, drifted
\end{minipage} & \begin{minipage}[b]{\linewidth}\raggedleft
O-Funnel, drifted
\end{minipage} \\*
\midrule\noalign{}
\endhead
\bottomrule\noalign{}
\endlastfoot
pmid & 1.000 & 1.000 & 0.000 & 1.000 \\*
title & 1.000 & 1.000 & 0.000 & 1.000 \\*
issn & 1.000 & 1.000 & 0.000 & 1.000 \\*
language & 1.000 & 1.000 & 0.000 & 0.999 \\*
pub\_year & 1.000 & 1.000 & 0.000 & 1.000 \\*
doi & 1.000 & 1.000 & 1.000 & 1.000 \\*
\textbf{overall} & \textbf{1.000} & \textbf{1.000} & \textbf{0.199} &
\textbf{1.000} \\*
\end{longtable}

On pristine data O-Funnel matches both the hand-written parser and the
regexes exactly, while also verifying that every capture is complete
(the oracle passed on 100\% of records, pristine and drifted). After the
drift the regexes lose every renamed field and keep only the DOI, whose
tags did not change. O-Funnel keeps all six. The renamed PMID and ISSN
are resolved by the synonym rung, \texttt{TitleOfArticle} and
\texttt{Lang} by the key-spelling rung, and the year by the key rung
once \texttt{within="PubDate"} is matched to \texttt{PublicationDate} by
key spelling. The DOI is found by shape in both conditions. The
remaining 0.1\% error on \texttt{language} is instructive: in 38 records
an \texttt{\textless{}OtherAbstract\ Language="eng"\textgreater{}}
element carries an \emph{attribute} literally named \texttt{Language}.
Once the article's \texttt{Language} element is renamed, that attribute
becomes the only exact key match and outranks the renamed element.
Declaring \texttt{within="Article"} removes it; we report the unscoped
requirement as written.

\textbf{What the text-bearing element view contributes.} Running the
identical script with O-Funnel 0.1.0, which lacks the value view of
Section 4.1, gives pristine F1 of 0.000 for pmid, 0.995 for title, 0.959
for issn, 1.000 for language and pub\_year, and 0.000 for doi: 0.820
overall, and 0.820 after drift. The PMID and DOI sit in
attribute-bearing elements that the earlier version captured correctly
but could not resolve as values, and the title was missed exactly on the
records with inline markup. The ISSN figure is partly luck: unable to
reach the attribute-bearing \texttt{ISSN} element, the earlier version
fell back through the key-spelling rung to the neighboring
\texttt{ISSNLinking} element, whose value often but not always equals
the journal ISSN (in the recent file, only 55\% of the time). A
plausible neighbor quietly standing in for the requested field is
exactly the failure the value view removes. Without the synonym map (a
second ablation), drifted F1 is 0.000 for pmid and 0.959 for issn (again
through the \texttt{ISSNLinking} neighbor), 0.893 overall: the
key-spelling rung alone cannot bridge \texttt{PubMedIdentifier} to
\texttt{pmid}, which is exactly the gap the ontology hook exists to
fill.

\textbf{Throughput.} On these real records (much larger than the
synthetic ones, about 50 elements each) O-Funnel processes 144 records
per second including capture, the completeness oracle and six
requirements, on one core.

\subsection{Constructed challenge
suites}\label{constructed-challenge-suites}

The next two suites are built on purpose to exercise the failure modes
of byte-level extraction. They show behavior under controlled
conditions; they are not a sample of real-world difficulty.

\textbf{Perturbation suite.} 13 base records rendered under nine drift
categories give 117 documents with four fields each: pristine XML, keys
renamed to registered synonyms, camelCase keys, JSON, JSON with renamed
keys, CSV, a long-form date, a same-shaped decoy placed before the true
value, and opaque keys (\texttt{c1} to \texttt{c4}).

\begin{longtable}[]{@{}lrrr@{}}
\caption{Perturbation suite, 117 documents, F1 per drift
category.}\tabularnewline
\toprule\noalign{}
category & regex & O-Funnel & O-Funnel + learned \\*
\midrule\noalign{}
\endfirsthead
\toprule\noalign{}
category & regex & O-Funnel & O-Funnel + learned \\*
\midrule\noalign{}
\endhead
\bottomrule\noalign{}
\endlastfoot
pristine XML & 0.86 & 1.00 & 1.00 \\*
renamed (synonym) & 0.86 & 0.86 & 0.86 \\*
camelCase & 0.86 & 1.00 & 1.00 \\*
JSON & 0.86 & 1.00 & 1.00 \\*
JSON renamed & 0.86 & 0.86 & 0.86 \\*
CSV & 0.86 & 1.00 & 1.00 \\*
long-form date & 0.67 & 1.00 & 1.00 \\*
decoy first & 0.40 & 1.00 & 1.00 \\*
opaque keys & 0.86 & 0.00 & 0.67 \\*
\textbf{overall} & \textbf{0.80} & \textbf{0.91} & \textbf{0.94} \\*
\end{longtable}

O-Funnel matches or beats the regexes everywhere except opaque keys.
There, before learning, it declines to guess (0.00) and the regexes'
value patterns win; after the funnel learns the opaque aliases from the
residue it recovers to 0.67. The decoy category shows the two-space
separation most clearly: the regexes, forced to locate by position, grab
the distractor (0.40), while O-Funnel locates structurally and confirms
by shape (1.00). In the two ``renamed'' rows O-Funnel ties the regexes:
their fourth field is free text with no shape, renamed beyond the
key-spelling threshold, so both systems recover the other three fields
and miss the fourth.

\textbf{Hard scenarios.} 120 documents (15 records by 8 classes), each
class isolating one way byte regexes fail.

\begin{longtable}[]{@{}lrrr@{}}
\caption{Hard scenarios, 120 documents, F1 per failure
class.}\tabularnewline
\toprule\noalign{}
class & regex & O-Funnel & O-Funnel + scavenger \\*
\midrule\noalign{}
\endfirsthead
\toprule\noalign{}
class & regex & O-Funnel & O-Funnel + scavenger \\*
\midrule\noalign{}
\endhead
\bottomrule\noalign{}
\endlastfoot
free-text fields & 0.00 & 1.00 & 1.00 \\*
multi-date attribution & 0.00 & 1.00 & 1.00 \\*
decoy identifier & 0.00 & 1.00 & 1.00 \\*
date, long format & 0.00 & 1.00 & 1.00 \\*
date, US format & 0.00 & 1.00 & 1.00 \\*
date decoy in a title & 0.00 & 1.00 & 1.00 \\*
JSON free text & 0.00 & 1.00 & 1.00 \\*
fully opaque keys & 1.00 & 0.00 & 1.00 \\*
\textbf{overall} & \textbf{0.43} & \textbf{0.84} & \textbf{1.00} \\*
\end{longtable}

The regexes score zero on each class designed against them and 1.00 only
where a value pattern alone suffices. O-Funnel resolves the seven
structural classes; the scavenger then recovers the opaque-key class
while its uniqueness and margin guards keep it from grabbing decoys
elsewhere.

\subsection{Self-improvement from
residue}\label{self-improvement-from-residue}

A corpus of 24 records has its identifier and decision keys drifted to
opaque tokens (\texttt{f7}, \texttt{disp}). We measure recall, run the
funnel, promote its safe proposals, and measure again.

\begin{longtable}[]{@{}
  >{\raggedright\arraybackslash}p{(\linewidth - 8\tabcolsep) * \real{0.1579}}
  >{\raggedleft\arraybackslash}p{(\linewidth - 8\tabcolsep) * \real{0.2105}}
  >{\raggedleft\arraybackslash}p{(\linewidth - 8\tabcolsep) * \real{0.2105}}
  >{\raggedleft\arraybackslash}p{(\linewidth - 8\tabcolsep) * \real{0.2105}}
  >{\raggedleft\arraybackslash}p{(\linewidth - 8\tabcolsep) * \real{0.2105}}@{}}
\caption{Funnel gain on a corpus with keys drifted to opaque tokens,
recall before and after promotion.}\tabularnewline
\toprule\noalign{}
\begin{minipage}[b]{\linewidth}\raggedright
field
\end{minipage} & \begin{minipage}[b]{\linewidth}\raggedleft
no exemplars: before
\end{minipage} & \begin{minipage}[b]{\linewidth}\raggedleft
after
\end{minipage} & \begin{minipage}[b]{\linewidth}\raggedleft
with 2 exemplars: before
\end{minipage} & \begin{minipage}[b]{\linewidth}\raggedleft
after
\end{minipage} \\*
\midrule\noalign{}
\endfirsthead
\toprule\noalign{}
\begin{minipage}[b]{\linewidth}\raggedright
field
\end{minipage} & \begin{minipage}[b]{\linewidth}\raggedleft
no exemplars: before
\end{minipage} & \begin{minipage}[b]{\linewidth}\raggedleft
after
\end{minipage} & \begin{minipage}[b]{\linewidth}\raggedleft
with 2 exemplars: before
\end{minipage} & \begin{minipage}[b]{\linewidth}\raggedleft
after
\end{minipage} \\*
\midrule\noalign{}
\endhead
\bottomrule\noalign{}
\endlastfoot
identifier (shaped) & 0.00 & 1.00 & 0.00 & 1.00 \\*
decision (free text) & 0.00 & 0.00 & 0.00 & 1.00 \\*
\textbf{overall} & \textbf{0.00} & \textbf{0.50} & \textbf{0.00} &
\textbf{1.00} \\*
\end{longtable}

Without exemplars the funnel recovers the shape-identifiable identifier
and, correctly, nothing for the free-text decision, which has no
structural signal. Given two example values, the decision's value
population alone identifies \texttt{disp}, and it is promoted too. The
contest check is what makes this safe. In a third variant the decision
values are single words, like the neighboring \texttt{STATUS} value
(\texttt{Final}): both \texttt{disp} and \texttt{STATUS} are proposed
for the decision, both are marked contested in all 24 records, and
neither is promoted, while the identifier alias still is. The funnel
prefers leaving a field absent to learning a wrong alias.

\subsection{Schema matching sanity check
(Valentine)}\label{schema-matching-sanity-check-valentine}

To check that the key-spelling and value-profile machinery is sound on
its own, we use it as a schema matcher on the public Valentine benchmark
{[}13{]} and compare it with the classical matchers bundled in the
Valentine framework, run with default settings. The metric is recall at
ground truth (rank predicted column pairs, keep the top
\textbar gold\textbar, count gold pairs recovered), micro-averaged per
category. The 36 instances are held out: a seeded sample disjoint from
the sample we used during development, covering the three categories
with enough instances left over.

\begin{longtable}[]{@{}
  >{\raggedright\arraybackslash}p{(\linewidth - 14\tabcolsep) * \real{0.1944}}
  >{\raggedleft\arraybackslash}p{(\linewidth - 14\tabcolsep) * \real{0.1111}}
  >{\raggedleft\arraybackslash}p{(\linewidth - 14\tabcolsep) * \real{0.1111}}
  >{\raggedleft\arraybackslash}p{(\linewidth - 14\tabcolsep) * \real{0.1111}}
  >{\raggedleft\arraybackslash}p{(\linewidth - 14\tabcolsep) * \real{0.1111}}
  >{\raggedleft\arraybackslash}p{(\linewidth - 14\tabcolsep) * \real{0.1111}}
  >{\raggedleft\arraybackslash}p{(\linewidth - 14\tabcolsep) * \real{0.1111}}
  >{\raggedleft\arraybackslash}p{(\linewidth - 14\tabcolsep) * \real{0.1389}}@{}}
\caption{Valentine held-out subset (36 instances), recall at ground
truth. O-F: O-Funnel (names only, or fused with the value profile);
COMA-S and COMA-I: COMA schema-only and with instances; Distr.:
distribution-based; SimFlood: Similarity Flooding.}\tabularnewline
\toprule\noalign{}
\begin{minipage}[b]{\linewidth}\raggedright
category
\end{minipage} & \begin{minipage}[b]{\linewidth}\raggedleft
O-F names
\end{minipage} & \begin{minipage}[b]{\linewidth}\raggedleft
O-F fused
\end{minipage} & \begin{minipage}[b]{\linewidth}\raggedleft
COMA-S
\end{minipage} & \begin{minipage}[b]{\linewidth}\raggedleft
COMA-I
\end{minipage} & \begin{minipage}[b]{\linewidth}\raggedleft
Cupid
\end{minipage} & \begin{minipage}[b]{\linewidth}\raggedleft
Distr.
\end{minipage} & \begin{minipage}[b]{\linewidth}\raggedleft
SimFlood
\end{minipage} \\*
\midrule\noalign{}
\endfirsthead
\toprule\noalign{}
\begin{minipage}[b]{\linewidth}\raggedright
category
\end{minipage} & \begin{minipage}[b]{\linewidth}\raggedleft
O-F names
\end{minipage} & \begin{minipage}[b]{\linewidth}\raggedleft
O-F fused
\end{minipage} & \begin{minipage}[b]{\linewidth}\raggedleft
COMA-S
\end{minipage} & \begin{minipage}[b]{\linewidth}\raggedleft
COMA-I
\end{minipage} & \begin{minipage}[b]{\linewidth}\raggedleft
Cupid
\end{minipage} & \begin{minipage}[b]{\linewidth}\raggedleft
Distr.
\end{minipage} & \begin{minipage}[b]{\linewidth}\raggedleft
SimFlood
\end{minipage} \\*
\midrule\noalign{}
\endhead
\bottomrule\noalign{}
\endlastfoot
ChEMBL & 0.60 & 0.67 & 0.53 & \textbf{0.76} & 0.33 & 0.16 & 0.55 \\*
OpenData & \textbf{0.72} & 0.69 & 0.69 & 0.55 & 0.31 & 0.18 & 0.46 \\*
TPC-DI & 0.70 & 0.75 & 0.80 & \textbf{0.86} & 0.19 & 0.52 & 0.58 \\*
\textbf{overall} & 0.69 & \textbf{0.70} & 0.67 & 0.66 & 0.29 & 0.23 &
0.50 \\*
\end{longtable}

Overall O-Funnel is on par with COMA and ahead of the other classical
matchers, with no training and no external resources beyond a small
abbreviation table. Per category the picture is mixed and we report it
as such: COMA with instances wins ChEMBL and TPC-DI, where column names
are weak and value distributions carry the signal, and O-Funnel wins
OpenData. Fusing the value profile helps where names diverge and costs a
little where names already carry the signal. The subset is small and
each matcher ran once, so these numbers are indicative. We did not
compare against learned or language-model matchers {[}27, 28, 29{]}; the
point here is that the localization machinery is sound, not that
O-Funnel is the best schema matcher.

\subsection{An out-of-scope probe
(SWDE)}\label{an-out-of-scope-probe-swde}

On the first 500 rows of a SWDE-derived web-extraction set {[}5, 31{]},
where each row pairs a movie-page HTML slice with an attribute name and
its gold value, O-Funnel recovers 0.2\% of values from the attribute
name alone. A value regex does better where a shape exists (51\% for
rating, 8\% for year). The reason is diagnostic: on these pages the
field name lives in a separate label element next to the value
(\texttt{\textless{}td\textgreater{}Director\textless{}/td\textgreater{}\textless{}td\textgreater{}...\textless{}/td\textgreater{}}),
a relation none of O-Funnel's rungs model. The value is present and
captured; what is missing is a label-adjacency rung. We report this to
mark the boundary of the current design.

\subsection{Throughput}\label{throughput}

\begin{longtable}[]{@{}lrr@{}}
\caption{Throughput on small synthetic documents, one
core.}\tabularnewline
\toprule\noalign{}
system & ms per document & documents per second \\*
\midrule\noalign{}
\endfirsthead
\toprule\noalign{}
system & ms per document & documents per second \\*
\midrule\noalign{}
\endhead
\bottomrule\noalign{}
\endlastfoot
byte regex & 0.008 & about 130,000 \\*
O-Funnel, pristine (fast path) & 0.295 & about 3,400 \\*
O-Funnel, every field drifted & 0.357 & about 2,800 \\*
O-Funnel, pristine + scavenger & 0.310 & about 3,200 \\*
\end{longtable}

O-Funnel is about 40 times slower than a bare regex, yet still processes
about 3,400 small documents per second on one core, ample for batch and
streaming ingestion. Time splits into capture 23\%, the completeness
oracle 18\% and resolution 59\%. A fully drifted document, where every
field is found by the non-key rungs, costs only 1.4 times the pristine
fast path, because key similarities are memoized across a feed.

\subsection{What each component
contributes}\label{what-each-component-contributes}

Table 9 maps each component to the scenario it resolves and the measured
effect of turning it on.

\begin{longtable}[]{@{}
  >{\raggedright\arraybackslash}p{(\linewidth - 4\tabcolsep) * \real{0.2857}}
  >{\raggedright\arraybackslash}p{(\linewidth - 4\tabcolsep) * \real{0.3061}}
  >{\raggedright\arraybackslash}p{(\linewidth - 4\tabcolsep) * \real{0.4082}}@{}}
\caption{Components, the scenarios they resolve, and their measured
effect.}\tabularnewline
\toprule\noalign{}
\begin{minipage}[b]{\linewidth}\raggedright
component
\end{minipage} & \begin{minipage}[b]{\linewidth}\raggedright
scenario it resolves
\end{minipage} & \begin{minipage}[b]{\linewidth}\raggedright
effect
\end{minipage} \\*
\midrule\noalign{}
\endfirsthead
\toprule\noalign{}
\begin{minipage}[b]{\linewidth}\raggedright
component
\end{minipage} & \begin{minipage}[b]{\linewidth}\raggedright
scenario it resolves
\end{minipage} & \begin{minipage}[b]{\linewidth}\raggedright
effect
\end{minipage} \\*
\midrule\noalign{}
\endhead
\bottomrule\noalign{}
\endlastfoot
structural localization + shape confirmation & decoy before the real
value & 0.40 to 1.00 (perturbation, decoy first) \\*
value normalizers & long-form and US dates & 0.00 to 1.00 (hard
scenarios, both date classes) \\*
text-bearing element view & attribute-bearing and mixed-content elements
& 0.820 to 1.000 (PubMed, pristine) \\*
synonym rung & renamed identifiers with no shared spelling & 0.893 to
1.000 (PubMed, drifted) \\*
key-spelling rung & renamed keys with shared words or abbreviations &
camelCase 0.86 to 1.00; PubMed title, language, year kept under drift \\*
scavenger & fully opaque keys & 0.00 to 1.00 (hard scenarios, opaque
keys) \\*
residue funnel & opaque keys across a corpus & 0.00 to 0.67
(perturbation, opaque keys) \\*
value profile in the funnel & renamed free-text field & 0.00 to 1.00
(Section 6.4) \\*
\end{longtable}

\section{Discussion}\label{discussion}

\textbf{What the two-space separation buys.} Every result above traces
back to one design decision: never match against bytes. Localization is
structural, value description is a witness, and the two are fused rather
than packed into one pattern. Drift in one dimension (key name, value
encoding, format, decoy) no longer breaks the other, because different
machinery handles each.

\textbf{Practical implications.} Much data-engineering effort goes not
into analysis but into keeping ingestion alive: parsers that were
correct last quarter break when a source renames a key, changes an
encoding, switches format or adds a lookalike field, and the usual
failure is a silent null discovered downstream. O-Funnel changes three
things in such systems. Silent failure becomes loud failure: a capture
that loses data is rejected, and a missing field comes with a reason.
Schema drift becomes a degradation rather than an incident: a renamed
field is recovered by the non-key rungs and, once promoted, resolved
cheaply again. And every value is auditable: its method, confidence and
provenance allow a review queue built around exactly the uncertain
cases. The funnel's idea, using what an extractor failed to claim as the
signal for what to learn next, applies to any extractor that can name
its residue.

\textbf{Limitations.} O-Funnel reads structure, not meaning: it has no
language model and cannot infer a field from prose semantics. It assumes
each document is individually well formed in a supported format; a
malformed XML or JSON document fails capture and must be handled
upstream, for example by quarantine. It does not model label adjacency
(Section 6.6) and does not learn per-site templates, so presentational
HTML is largely out of scope. PDF and scanned documents need an OCR and
layout front end first. Our documents are small (a handful to about
fifty elements); we have not studied very wide schemas with hundreds of
candidate fields, where ambiguity and the \(O(N \cdot R)\) cost both
grow. The confidence constants are hand-set and were not tuned or
studied for sensitivity.

\textbf{Threats to validity.} The constructed suites are ours and probe
the failure modes we anticipated; we release their generators so others
can extend them. The real-corpus drift is simulated by renaming
elements, a common but not the only form of real drift. The Valentine
subset is small and ran once per matcher. We mitigate these by reporting
per-category results, showing where O-Funnel loses, and publishing every
script.

\section{Conclusion}\label{conclusion}

Brittle extraction is usually blamed on brittle patterns, but the
pattern is the symptom. The cause is matching a structured document
through an unstructured byte view, which forces one expression to be
fluent in two spaces at once. O-Funnel separates them: it captures any
document losslessly into one structural vocabulary with a completeness
guarantee, declares each requirement in that vocabulary, and resolves it
by fusing independent evidence, with unclaimed data feeding a
self-improvement loop. On real PubMed records it matches a hand-written
parser and keeps doing so after the schema drifts under it, where the
parser's regexes collapse; on constructed scenarios it resolves every
failure class designed against regexes; and it holds its own as a schema
matcher without training. The negative result on adjacency-labelled web
pages marks the boundary of the current design and names the next rung
to build.

\section{Code and Data Availability}\label{code-and-data-availability}

O-Funnel is released under the MIT license at
https://github.com/osamaa-mustafa/ofunnel and on PyPI
(\texttt{pip\ install\ ofunnel}, version 0.1.1). The repository's
\texttt{benchmarks/} directory contains the script behind every number
in Section 6. The PubMed baseline {[}30{]}, the Valentine datasets
{[}13{]} and the SWDE-derived rows {[}31{]} are public.

\end{document}